\documentclass[prd,onecolumn,nofootinbib,showkeys]{revtex4}
\usepackage{epsfig}
\newcommand\gothfamily{\usefont{U}{ygoth}{m}{n}}
\DeclareTextFontCommand{\textgoth}{\gothfamily}

\begin{document}

\title{RADIAL MOTION INTO AN EINSTEIN-ROSEN BRIDGE}

\author{{\bf Nikodem J. Pop\l awski}}

\affiliation{Department of Physics, Indiana University, Swain Hall West, 727 East Third Street, Bloomington, IN 47405, USA}
\email{nipoplaw@indiana.edu}

\noindent
{\em Physics Letters B}\\
Vol. {\bf 687}, Nos. 2-3 (2010) pp. 110--113\\
\copyright\,Elsevier B. V.
\vspace{0.4in}

\begin{abstract}
We consider the radial geodesic motion of a massive particle into a black hole in isotropic coordinates, which represents the exterior region of an Einstein-Rosen bridge (wormhole).
The particle enters the interior region, which is regular and physically equivalent to the asymptotically flat exterior of a white hole, and the particle's proper time extends to infinity.
Since the radial motion into a wormhole after passing the event horizon is physically different from the motion into a Schwarzschild black hole, Einstein-Rosen and Schwarzschild black holes are different, physical realizations of general relativity.
Yet for distant observers, both solutions are indistinguishable.
We show that timelike geodesics in the field of a wormhole are complete because the expansion scalar in the Raychaudhuri equation has a discontinuity at the horizon, and because the Einstein-Rosen bridge is represented by the Kruskal diagram with Rindler's elliptic identification of the two antipodal future event horizons.
These results suggest that observed astrophysical black holes may be Einstein-Rosen bridges, each with a new universe inside that formed simultaneously with the black hole.
Accordingly, our own Universe may be the interior of a black hole existing inside another universe.
\end{abstract}


\maketitle

\section{Isotropic coordinates}

The interval of the static, spherically symmetric, gravitational field in vacuum, expressed in isotropic coordinates, was found by Weyl \cite{Weyl}:
\begin{equation}
ds^2=\frac{(1-r_g/(4r))^2}{(1+r_g/(4r))^2}c^2 dt^2-(1+r_g/(4r))^4(dr^2+r^2 d\Omega^2),
\label{interv1}
\end{equation}
where $0\le r<\infty$ is the radial coordinate, $d\Omega$ is the element of the solid angle, and $r_g=2GM/c^2$ is the Schwarzschild radius.
This metric does not change its form under the coordinate transformation:
\begin{equation}
r\rightarrow r'=\frac{r_g^2}{16r},
\label{transf}
\end{equation}
and is Galilean for $r\rightarrow\infty$.
Therefore it is also Galilean for $r\rightarrow 0$, describing an Einstein-Rosen bridge (wormhole): two Schwarzschild solutions to the Einstein field equations (a black hole and white hole) connected at the singular ($\textrm{det}\,g_{\mu\nu}=0$) surface $r=r_g/4$ (common event horizon) \cite{ER,FW}.
The nonzero components of the Riemann curvature tensor for this metric are given by
\begin{eqnarray}
& & R^{0\theta}_{\phantom{0\theta}0\theta}=R^{0\phi}_{\phantom{0\phi}0\phi}=R^{r\theta}_{\phantom{r\theta}r\theta}=R^{r\phi}_{\phantom{r\phi}r\phi}=\frac{r_g}{2r^3(1+r_g/(4r))^6}, \nonumber \\
& & R^{0r}_{\phantom{0r}0r}=R^{\theta\phi}_{\phantom{\theta\phi}\theta\phi}=-\frac{r_g}{r^3(1+r_g/(4r))^6},
\label{curv}
\end{eqnarray}
so the Kretschmann scalar is finite everywhere: $R_{\mu\nu\rho\sigma}R^{\mu\nu\rho\sigma}=12r_g^2 r^{-6}(1+r_g/(4r))^{-12}$, going to zero as $r\rightarrow\infty$ and $r\rightarrow 0$.
The Einstein-Rosen metric for $r>r_g/4$ describes the exterior sheet of a Schwarzschild black hole (the transformation of the radial coordinate $r\rightarrow r_S=r(1+r_g/(4r))^2$ brings the interval (\ref{interv1}) into the standard Schwarzschild form \cite{Schw}).
The spacetime given by the metric (\ref{interv1}) for $r<r_g/4$ is regarded by an observer at $r>r_g/4$ as the interior of a black hole.
Because of the invariance of the metric (\ref{interv1}) under the transformation (\ref{transf}), this interior is an image of the other exterior sheet.
This situation is analogous to the method of image charges for spheres in electrostatics, where the interaction between an electric charge situated at a distance $r$ from the center of a conducting sphere of radius $R<r$ is equivalent to the interaction of the same charge with a charge of the opposite sign situated inside this sphere at a distance $R^2/r$ from its center \cite{LL8}.
The radius $R$ corresponds to the Schwarzschild surface in isotropic coordinates, $r=r_g/4$.

\section{Radial motion}

Consider a massive particle moving radially in the gravitational field described by the metric (\ref{interv1}).
For brevity, we use
\begin{equation}
h=g_{00}=\frac{(1-r_g/(4r))^2}{(1+r_g/(4r))^2},\,\,\,\,f=-g_{rr}=(1+r_g/(4r))^4.
\label{not}
\end{equation}
The motion of the particle is given by the radial geodesic equations.
If the particle is at rest at $r=r_0$, then these equations are
\begin{eqnarray}
& & \frac{dt}{d\tau}=u^0=\sqrt{h_0}/h, \label{mot1} \\
& & \frac{dr}{cd\tau}=u^r=\epsilon(h_0 h^{-1} f^{-1}-f^{-1})^{1/2},
\label{mot2}
\end{eqnarray}
where $\tau$ is the proper time of the particle, $h_0=h|_{r=r_0}$, and $\epsilon=-1\,(+1)$ for an infalling (outgoing) motion.
Consider a particle falling into a black hole, $\epsilon=-1$.
As $r\rightarrow r_g/4$, $h$ goes to zero and $f\rightarrow 16$, so both $u^0$ and $u^r$ become infinite.
Even if the initial motion were not purely radial, the components $u^0,\,u^r$ would still become infinite at $r=r_g/4$, with $u^\theta,\,u^\phi$ remaining finite.
Therefore, each motion of a massive particle becomes effectively radial at the surface $r=r_g/4$.

A distant observer situated in a nearly Galilean spacetime measures the velocity of the infalling particle as
\begin{equation}
v_d=\frac{dr}{dt}=c\frac{u^r}{u^0}=-c\frac{(h_0 f^{-1}h-f^{-1}h^2)^{1/2}}{\sqrt{h_0}}.
\label{mot3}
\end{equation}
As $r\rightarrow r_g/4$, $v_d$ goes to zero.
Writing $r=r_g/4+\xi$, where $0<\xi\ll r_g$, gives $v_d\approx -c\xi/(2r_g)$ and thus $r-r_g/4\sim\mbox{exp}(-ct/(2r_g))$, so the particle reaches the surface of a black hole $r=r_g/4$ after an infinite time $t$.
This surface is an event horizon for a distant observer, as it is for the standard Schwarzschild metric \cite{Lem}.
The proper time $\Delta\tau$ of the particle for moving radially from $r=r_0$ to $r=r_g/4$ is finite, which can be shown by considering $r_0=r_g/4+\xi$:
\begin{equation}
c\Delta\tau=\int_{r_g/4}^{r_g/4+\xi}dr/u^r\approx r_g.
\label{mot4}
\end{equation}
After reaching the surface $r=r_g/4$, the particle continues moving; its radial coordinate $r$ decreases to $r_1=r_g^2/(16r_0)\approx r_g/4-\xi$ (at which $u^r=0$) in a proper time $\Delta\tau=r_g/c$.
The radial motion of a massive particle (in terms of the proper time) in the spacetime (\ref{interv1}) for $r\le r_g/4$ is the image (in the sense of the method of image charges for spheres in electrostatics) of the particle's motion for $r\ge r_g/4$.
Applying the transformation (\ref{transf}) to Eqs. (\ref{mot1}) and (\ref{mot2}) in the region $r\le r_g/4$ gives
\begin{eqnarray}
& & \frac{dt}{d\tau}=\sqrt{h_0}/h, \label{motin1} \\
& & \frac{dr'}{cd\tau}=-\epsilon(h_0 h^{-1} f^{-1}-f^{-1})^{1/2},
\label{motin2}
\end{eqnarray}
where $h=(1-r_g/(4r'))^2/(1+r_g/(4r'))^2$ and $f=(1+r_g/(4r'))^4$.
An infalling radial geodesic motion inside a black hole appears, in terms of the new radial coordinate $r'$, as an outgoing motion from a white hole (the time reversal of a black hole).

The local velocity of the particle $v_l$, measured in terms of the proper time, as determined by static clocks synchronized along the trajectory of the particle, is related to $u^0$ by $u^0=(h(1-v_l^2/c^2))^{-1/2}$ \cite{LL2}.
As the particle moves from $r=r_0$ to $r=r_g/4$, $v_l$ increases from zero to $c$, and as the particle moves from $r=r_g/4$ to $r=r_1$, $v_l$ decreases to zero.
In a Schwarzschild field, $v_l$ exceeds $c$ inside a black hole, which does not violate Einstein's theory of relativity because the interior of a Schwarzschild black hole is not static and neither can be clocks synchronized along the trajectory of the particle.

\begin{figure}[ht]
\centering
\includegraphics[width=4in]{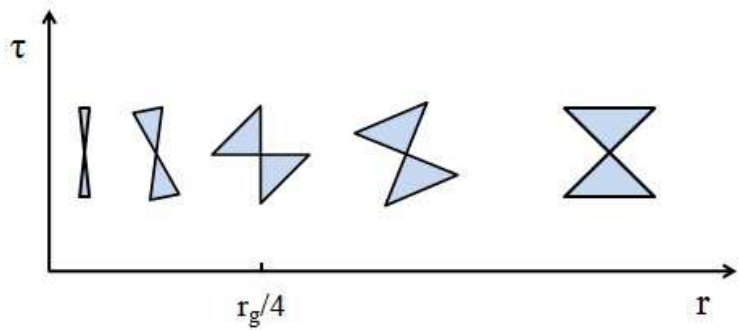}
\caption{Light cones in the gravitational field represented by the metric (\ref{interv2}).}
\label{fig1}
\end{figure}

Consider radially moving geodesic clocks that are falling into a black hole from the isotropic radius $r_0$, and synchronized such that the time of each clock at the instant of release equals the proper time $\tau_0$ of a clock at rest that remains fixed at $r_0$.
The time at any event is taken to be equal to the proper time on the radially falling clock that is coincident with this event, following the procedure due to Gautreau and Hoffmann \cite{GH}.
Equations (\ref{mot1}) and (\ref{mot2}) give (for $\epsilon=-1$)
\begin{equation}
cd\tau=\sqrt{h_0}cdt+\sqrt{f(h_0-h)/h}dr,
\label{mot5}
\end{equation}
which, using $\tau_0=\sqrt{h_0}t_0$, integrates to
\begin{equation}
c\tau=\sqrt{h_0}ct+\int_{r_0}^{r}\sqrt{f(h_0-h)/h}dr,
\label{mot6}
\end{equation}
giving the transformation between the coordinates $(r,t)$ and $(r,\tau)$.
In terms of $\tau$, the metric (\ref{interv1}) becomes
\begin{equation}
ds^2=h(cd\tau-\sqrt{f(h_0-h)/h}dr)^2/h_0-fdr^2-fr^2d\Omega^2.
\label{interv2}
\end{equation}
Radial null geodesics are given by $ds=0$ and $d\Omega=0$:
\begin{equation}
\frac{cd\tau}{dr}=\sqrt{fh_0/h}(\sqrt{1-h/h_0}\pm1).
\label{nul}
\end{equation}
The plus (minus) sign corresponds to an outgoing (infalling) null geodesic.
For $r\rightarrow\infty$, the spacetime is Galilean and $cd\tau/dr=\pm1$.
For $r=r_g/4$, $cd\tau/dr=\infty$ for the outgoing null geodesic and $0$ for the infalling one, as shown in Fig.~\ref{fig1}.
For $r=0$, $cd\tau/dr=\pm\infty$; however, in terms of the new radial coordinate $r'$ (\ref{transf}), we obtain $cd\tau/dr'=\pm1$ (Galilean spacetime).
Massive particles can move in both radial directions, except at the unidirectional surface $r=r_g/4$, where only infalling geodesics (decreasing $r$) lie inside the light cone.

The singularity theorem of Penrose guarantees that, whenever matter satisfies reasonable energy conditions, some sort of geodesic incompleteness occurs inside a black hole \cite{Wald}.
The key component in this theorem is the Raychaudhuri equation \cite{Ray}, describing the time evolution of the expansion scalar for a timelike congruence.
The expansion scalar $\theta=u^i_{\phantom{i};i}$ measures the fractional rate at which a small volume of matter changes with respect to time as measured by a comoving observer.
The Raychaudhuri equation guarantees that any timelike geodesic inside a Schwarzschild black hole converges in a caustic ($\theta\rightarrow -\infty$) within a finite proper time.
For the infalling radial motion into the Einstein-Rosen bridge (\ref{interv1}) from infinity ($h_0=1$), Eq. (\ref{mot2}) yields
\begin{equation}
\theta=\frac{1}{\sqrt{hf^3}r^2}(\sqrt{hf^3}r^2 u^r)_{,r}=-\frac{3}{2}\sqrt{\frac{r_g}{r^3}}\frac{\textrm{sgn}(1-r_g/(4r))}{(1+r_g/(4r))^3}.
\label{theta}
\end{equation}
As the particle moves from infinity to $r=r_g/4$, $\theta$ decreases from $0$ to $-3/(2r_g)$.
At $r=r_g/4$, $\theta$ undergoes a discontinuity, jumping to $3/(2r_g)$.
As the particle moves from $r=r_g/4$ to $r=0$, $\theta$ decreases back to zero.
The discontinuity of the expansion scalar at the event horizon of a wormhole prevents $\theta$ from decreasing to $-\infty$, as it occurs inside a Schwarzschild black hole.
Therefore this discontinuity guarantees that timelike geodesics in the gravitational field of a wormhole are complete.

\section{Kruskal representation}

\begin{figure}[ht]
\centering
\includegraphics[width=4in]{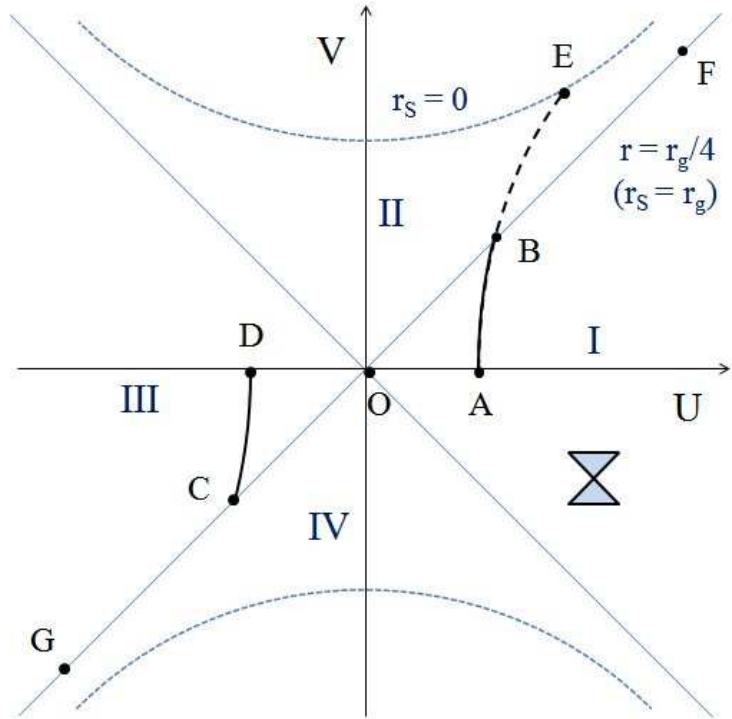}
\caption{Infalling radial geodesic motion of a massive particle into a Schwarzschild black hole ($ABE$) and an Einstein-Rosen black hole ($ABCD$) in the Kruskal coordinates. 
For the Einstein-Rosen black hole, the events on the segment $OF$ are identified with the events on the segment $OG$, e.g., points $B$ and $C$ represent the same spacetime event.}
\label{fig2}
\end{figure}

A maximal extension of the Schwarzschild metric \cite{Syn,KS} shows that a massive particle cannot travel in the spacetime of a Schwarzschild black hole from Kruskal's right-hand quadrant (exterior region I in Fig.~\ref{fig2}) to left-hand quadrant (exterior region III) without violating causality: the Schwarzschild bridge is not traversable \cite{FW,BMT}.
Such a particle either remains in region I or moves to the upper quadrant (interior region II), where it reaches the central singularity and its proper time ends.
Equation (\ref{mot3}) can be solved in the Schwarzschild coordinates, where $h=1-r_g/r_S$ and $f=(1-r_g/r_S)^{-1}$, in the parametric form \cite{Chan}:
\begin{eqnarray}
& & \frac{r_S}{r_{S0}}=\frac{1}{2}(1+\textrm{cos}\eta), \label{solut1} \\
& & \frac{ct}{r_g}=\textrm{ln}\frac{k+\textrm{tan}(\eta/2)}{k-\textrm{tan}(\eta/2)}+k\Bigl(\eta+\frac{r_{S0}}{2r_g}(\eta+\textrm{sin}\eta)\Bigr),
\label{solut2}
\end{eqnarray}
where $r_{S0}=r_0(1+r_g/(4r_0))^2$, $k=(r_{S0}/r_g-1)^{1/2}$, and the parameter $\eta$ goes from $0$ (the particle is at rest at $t=0$) to the value at the event horizon, $\eta_h=2\textrm{tan}^{-1}k$.
Equations (\ref{mot2}) and (\ref{solut1}) give the proper time $\tau$ as a monotonously increasing function of $\eta$:
\begin{equation}
\tau=\frac{1}{2c}\sqrt{\frac{r_{S0}^3}{r_g}}(\eta+\textrm{sin}\eta).
\label{prop}
\end{equation}
Transforming from the Schwarzschild coordinates $(r,t)$ to the Kruskal coordinates $(U,V)$ (in region I) \cite{KS},
\begin{eqnarray}
& & U=(r_S/r_g-1)^{1/2}e^{r_S/(2r_g)}\textrm{cosh}\frac{ct}{2r_g}, \label{Kr1} \\
& & V=(r_S/r_g-1)^{1/2}e^{r_S/(2r_g)}\textrm{sinh}\frac{ct}{2r_g},
\label{Kr2}
\end{eqnarray}
together with Eq. (\ref{mot3}), gives the derivative $dV/dU$:
\begin{equation}
\frac{dV}{dU}=\frac{k-\textrm{tanh}\frac{ct}{2r_g}\cdot(\frac{r_{S0}}{r_S}-1)^{1/2}}{k\textrm{tanh}\frac{ct}{2r_g}-(\frac{r_{S0}}{r_S}-1)^{1/2}}.
\label{der1}
\end{equation}
In the limit $\eta\rightarrow\eta_h$ we can write $\eta=\eta_h-\xi$, where $0<\xi\ll 1$, which gives $r_S/r_g-1\approx k\xi$, $\textrm{exp}(ct/(2r_g))\approx 2\textrm{cos}(\eta_h/2)\sqrt{k/\xi}\textrm{exp}[k(\eta_h+r_{S0}(\eta_h+\textrm{sin}\eta_h)/(2r_g))/2]$ and $(r_{S0}/r_S-1)^{1/2}\approx k-\xi/2$.
We obtain the values of $U,V$ for the particle at the event horizon:
\begin{equation}
U_h=V_h=\frac{\sqrt{e}k}{\sqrt{1+k^2}}e^{\frac{k}{2}\big((k^2+3)\textrm{tan}^{-1}k+k\big)},
\label{KrEH}
\end{equation}
and the value of the slope $dV/dU$ at the horizon:
\begin{equation}
\frac{dV}{dU}\Big|_h=\zeta=\textrm{coth}\Big[\frac{k}{2}\Big((k^2+3)\textrm{tan}^{-1}k+k\Big)\Big].
\label{der2}
\end{equation}

After the particle, which started moving from the radius $r_{S0}>r_g$ at $t=0$ (point $A$ in Fig.~\ref{fig2}), reaches the event horizon of a Schwarzschild black hole at $r_S=r_g,\,r=r_g/4$ (point $B$), it moves toward the central singularity at $r_S=0$ (point $E$) as $\eta$ (which is an increasing function of the particle's proper time $\tau$) goes to $\pi$.

The completeness of the particle's geodesic in the spacetime of the Einstein-Rosen bridge in the Kruskal coordinates can be explained if we note that both coordinate pairs $(-U,-V)$ and $(U,V)$ mathematically correspond to the same coordinates $r,t$.
Rindler suggested that both coordinate pairs $(-U,-V)$ and $(U,V)$ physically represent the same coordinates $r,t$, i.e. region III is identical with region I, and (interior) region IV is identical with region II; the Kruskal spacetime is elliptic \cite{Rin}.
In order to represent the Einstein-Rosen bridge in the Kruskal coordinates, we need to impose Rindler's elliptic identification of event antipodes {\em only} on the line $V=U$, i.e. only the two antipodal future event horizons are identical.
After the particle reaches the event horizon of an Einstein-Rosen black hole at point $B$, it moves from point $C$, which is {\em identical} with point $B$, to point $D$, which is related to point $A$ via the transformation (\ref{transf}) of the isotropic radial coordinate $r$.
As the particle moves from point $C$ to $D$, the proper time $\tau$ increases, while the coordinate time $t$ decreases (runs in the reverse direction with respect to observers in region I).
This reversion does not cause any problems, because it occurs after $t\rightarrow\infty$, so observers in region I never see it.
At infinity of region III, the proper time $\tau$ of the particle coincides with $-t$, so $-t$ has the meaning of the physical coordinate time of a white hole (region III).

The dependence of $r$ on $\eta$ in the radial geodesic motion of a massive particle into an Einstein-Rosen black hole is given by
\begin{equation}
r\Bigl(1+\frac{r_g}{4r}\Bigr)^2=\frac{1}{2}(1+\textrm{cos}\eta)r_0 \Bigl(1+\frac{r_g}{4r_0}\Bigr)^2.
\label{solut3}
\end{equation}
The segment $AB$ corresponds to the motion from $r=r_0$ ($\tau=0$, $dV/dU=\infty$) to $r=r_g/4$ ($\tau=\tau_h=\sqrt{r_{S0}^3/r_g}(\eta_h+\textrm{sin}\eta_h)/(2c)$, $dV/dU=\zeta$), and the segment $CD$ corresponds to the motion from $r=r_g/4$ ($\tau=\tau_h$, $dV/dU=\zeta$; the slope $dV/dU$ is continuous at the horizon as we go from point $B$ to $C$) to $r=r_1=r_g^2/(16r_0)$ ($\tau=2\tau_h$, $dV/dU=\infty$).
Therefore crossing the Einstein-Rosen bridge is possible if we regard Rindler's elliptic identification of the two antipodal future event horizons as physical.
The particle enters region III (with no possibility of coming back to region I), which has no curvature singularities and is mathematically equivalent to the asymptotically flat exterior of a white hole \cite{NN}.
The particle's proper time does not end in this region, but extends to infinity.

\section{Discussion}

The difference in the character of the radial motion inside the event horizon between an Einstein-Rosen black hole and a Schwarzschild black hole indicates that the two black hole solutions are physically different with regard to the nature of the interior sheet and equivalent with regard to the nature of the static exterior sheet (region I).
For an Einstein-Rosen black hole with Rindler's elliptic identification of the two antipodal future event horizons the interior (from the point of view of a distant observer; it really is the exterior region III) is static, while for a Schwarzschild black hole the interior (region II) is nonstatic.
Rindler's identification also guarantees that the Einstein-Rosen bridge is a stable solution to the gravitational field equations; without this identification the bridge would be unstable \cite{FW}.
Another difference between a Schwarzschild black hole and an Einstein-Rosen black hole comes from the integral of the time-time component of the gravitational energy-momentum pseudotensor (either Einstein or Landau-Lifshitz) over the interior hypersurface, i.e. the total energy of the system \cite{LL2,Schr}.
This energy equals the sensible physical value $r_gc^4/2G=Mc^2$ for an Einstein-Rosen black hole, but diverges for a Schwarzschild black hole.
Although the construction of the Einstein-Rosen bridge in the Kruskal coordinates through Rindler's elliptic identification of the two antipodal future event horizons may seem artificial, the necessity for Rindler's identification to avoid the particle's motion to the central singularity may be related to a singular behavior of the Kruskal metric in the Minkowski limit $r_g\rightarrow0$.
However, the Kruskal coordinates are advantageous in describing the Einstein-Rosen bridge because the surface $r=r_g/4$ is regular ($\textrm{det}\,g_{\mu\nu}<0$) in these coordinates.

The Schwarzschild black hole solution, singular at the center, does not exist in isotropic coordinates, while the Einstein-Rosen bridge (wormhole) geometry, regular everywhere, can be built in the Schwarzschild coordinates by gluing together two Schwarzschild exterior sheets at their common event horizon \cite{V}.
While the Schwarzschild metric is the spherically symmetric solution to the Einstein field equations in vacuum if we solve these equations using the Schwarzschild coordinates, the Einstein-Rosen bridge is the spherically symmetric solution to the Einstein field equations if we solve these equations using isotropic coordinates for a source which is vacuum everywhere except at the surface $r=r_g/4$.
It has been shown that the Einstein-Rosen bridge metric (\ref{interv1}) is not a solution of the vacuum Einstein equations but it requires the presence of a nonzero energy-momentum tensor source $T_{\mu\nu}$ that is divergent and violates the energy conditions at the throat of the wormhole \cite{V,Gue}.
This metric satisfies the Levi-Civita identity, $R^0_0=\frac{1}{\sqrt{g_{00}}}\nabla^2(\sqrt{g_{00}})$, where $\nabla^2$ is the Laplace-Beltrami operator \cite{Gue,Fra}.
This identity gives for (\ref{interv1}):
\begin{equation}
R^0_0=\frac{8\pi G}{c^4}\biggl(T^0_0-\frac{1}{2}T\biggr)=\frac{\delta(r-r_g/4)}{8|r-r_g/4|}=\frac{\delta\Bigl((r-r_g/4)^2\Bigr)}{4}.
\end{equation}
The curvature scalar at the throat of such a wormhole acquires a similar delta-function contribution.
It has been shown in \cite{Gue} that the delta-function matter source in the Einstein-Rosen bridge at $r=r_g/4$ is a lightlike brane self-consistently interacting with gravity.

Both black hole solutions are mathematically legitimate, and only experiment or observation can reveal the nature of the infalling radial motion of a particle into a physical black hole.
Since the two solutions are indistinguishable for distant observers, which can see only the exterior sheet, the nature of the interior of a physical black hole cannot be satisfactorily determined, unless an observer enters or resides in the interior region.
This condition would be satisfied if our Universe were the interior of a black hole existing in a bigger universe \cite{Pat}.
Because Einstein's general theory of relativity does not choose a time orientation, if a black hole can form from the gravitational collapse of matter through an event horizon in the future then the reverse process also possible.
Such a process would describe an exploding white hole: matter emerging from an event horizon in the past, like the expanding Universe.
Scenarios in which the Universe was born from the interior of an Einstein-Rosen black hole may avoid many of the problems of the standard Big-Bang cosmology and the black hole information-loss problem \cite{EB}.
These scenarios, involving gravitational collapse of a sphere of dust in isotropic coordinates, and generalization of the results of the present Letter to Schwarzschild-de Sitter and Kerr black holes will be the subjects of further study.

\end{document}